\documentclass[twocolumn,showpacs,showkeys,preprintnumbers,amsmath,amssymb,prb,floatfix]{revtex4}
\usepackage{graphicx}
\usepackage{dcolumn}
\usepackage{bm}
\usepackage{float}
\usepackage[utf8]{inputenc}
\usepackage[T1]{fontenc}

\begin{document}

 \title{High resolution Compton scattering as a Probe of the Fermi surface
 in the Iron-based superconductor LaO$_{1-x}$F$_{x}$FeAs}

 \author{Y. J. Wang$^1$, Hsin Lin$^1$,
B. Barbiellini$^1$, P.E. Mijnarends$^{1,2}$,\ \\ S. Kaprzyk$^{1,3}$,
W. Al-Sawai$^1$, R.S. Markiewicz$^1$ and A. Bansil$^1$}

 \affiliation{
$^1$Physics Department, Northeastern University, Boston, Massachusetts 02115, USA\\
$^2$Department of Radiation, Radionuclides $\&$ Reactors, Faculty of
Applied Sciences, Delft University of Technology, Delft, The
Netherlands \\
$^3$Academy of Mining and Metallurgy AGH, 30059 Krakow, Poland\\
}

 \date{\today}
 \keywords{Electric momentum density, Compton profile, Fermi surface, pnictide}
 \pacs{71.18.+y, 71.20.-b, 74.25.Jb 74.70.Dd}

 \begin{abstract}
We have carried out first principles all-electron calculations
of the (001)-projected 2D electron momentum density and the
directional Compton profiles along the [100], [001] and
[110] directions in the Fe-based superconductor LaOFeAs
within the framework of the local density approximation.
We identify Fermi surface features in the 2D electron momentum density and
the directional Compton profiles, and discuss issues related to the
observation of these features via Compton scattering experiments.
 \end{abstract}

\maketitle
\def\thesection{\arabic{section}}
\section{Introduction}
Since the discovery \cite{Kamihara} of superconductivity in a family
of iron-based superconductors (pnictides), there have been a large
number of studies of their electronic properties that have revealed
similarities between pnictides and cuprates. The theoretical prediction
of a striped antiferromagnetic spin-density-wave (SDW) ground state
\cite{Yildirim} was confirmed by neutron scattering \cite{Clarina}.
Superconductivity is found in LaOFeAs with either hole doping \cite{Hai}
or electron doping \cite{Kamihara,Chen}.
First principles calculations \cite{Boeri,Singh1}
find that the density of states (DOS) near the Fermi level ($E_F$)
is predominantly due to Fe-$d$ orbitals.
Owing to the approximate S$_4$ symmetry of the FeAs tetrahedra,
these Fe-$d$ orbitals split into lower lying
e$_{g}$ (d$_{x^{2}-y^{2}}$,d$_{3z^{2}-r^{2}}$) and
higher lying t$_{2g}$ states
(d$_{xy}$,d$_{yz}$,d$_{zx}$)\cite{Chao,Singh1}.
Theoretical calculations \cite{Boeri,Singh1} suggest that
superconductivity may not be caused by electron-phonon coupling.
Just as in the cuprates, the antiferromagnetic
instability, which is suppressed by doping, is one candidate to explain
unconventional superconductivity. Angle-resolved photoemission
spectroscopy (ARPES) experiments have recently been carried out in
BaFe$_2$As$_2$ \cite{Evtushinsky,HDing}, a related pnictide.
The superconducting gap in LaFeAsO$_{1-x}$F${_x}$ with $x \approx
10\%$ has been determined from the optical reflectance in the
far-infrared region.\cite{Chen}

X-ray scattering spectroscopy in the deeply inelastic (Compton) regime
provides a direct probe of the correlated many-body ground state in bulk
materials while avoiding the surface sensitivity of ARPES. The use of
modern synchrotron sources \cite{cooper} makes it possible to investigate
complex materials via the measurement of directional Compton
profiles\cite{Laukkanen}.

In this paper, we report first-principles computations of the 2D-projected
electron momentum density (2D-EMD) and Compton profiles (CPs) in the
iron-based superconductor LaOFeAs. We discuss Fermi surface (FS) images in
the 2D-EMD and its anisotropy defined by subtracting a smooth
isotropic function
from the spectrum. Our analysis of the CPs reveals that FS features
related to hole- as well as electron-pockets are more prominent in the CP
when the momentum transfer vector lies along the [100] rather than the
[110] direction.

\section{Methods}

\begin{figure*}[t]
\includegraphics[width=\hsize]{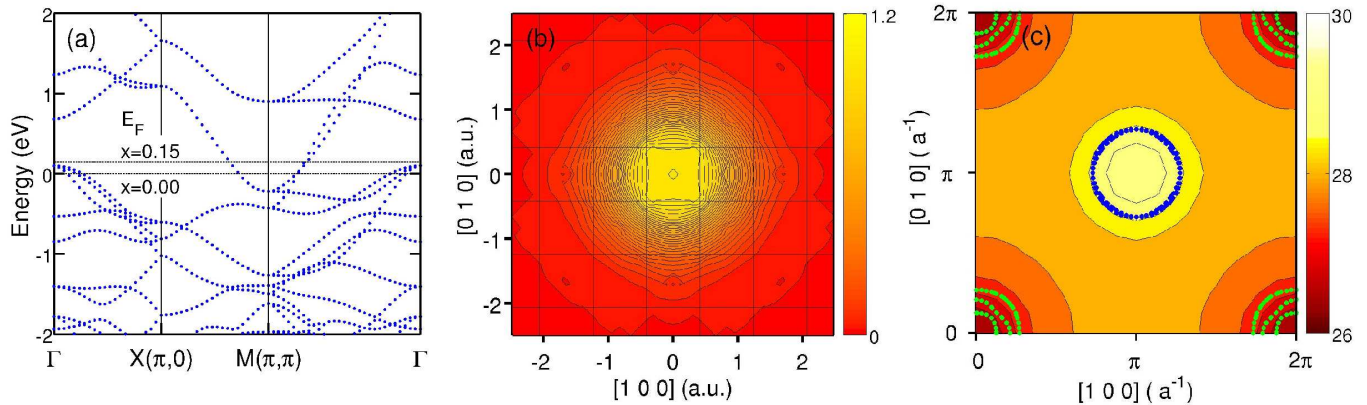}
\caption{(Color online)
(a) LDA band structure of LaOFeAs at $k_z$=0. (b) Theoretical 2D-EMD
$\rho^{2d}(p_x,p_y)$ of LaOFeAs projected onto the (001) plane, normalized
to $\rho^{2d}(0,0)$. (c) 2D-LCW folded momentum density. Computed
hole-like (green dots) and electron-like (blue dots) Fermi surfaces are
marked. Note momentum is given in (c) in units of $2\pi/a$, where
$a$=7.6052 a.u.}
\label{fig:EMD}
\end{figure*}

Our electronic structure calculations are based on the local density
approximation (LDA) of density functional theory. An all-electron fully
charge self-consistent semi-relativistic Korringa-Kohn-Rostoker (KKR)
method is used\cite{ABkkr}.
The compound LaO$_{1-x}$F$_x$FeAs has a simple tetragonal
structure (space-group P4/nmm). We used the experimental lattice
parameters\cite{Qiu2008} of LaO$_{0.87}$F$_{0.13}$FeAs in which no
spin-density-wave order was observed in neutron-scattering. A
non-spinpolarized calculation was performed and the magnetic structure was
neglected. Self-consistency was obtained for $x$=0 and the effects of
doping $x$ were treated within a rigid band model by shifting the Fermi
energy to accommodate the proper number of electrons. The convergence of
the crystal potential was approximately $10^{-4}$ Ry. The electron
momentum density (EMD) $\rho(p_x,p_y,p_z)$ was calculated on a momentum
mesh with step ($\Delta p_x,\Delta p_y,\Delta p_z) = (1/16a,1/16a,1/16c)2\pi$. The
total number of points was $14.58\times10^{6}$ within a sphere of radius
12.8 a.u. in momentum space. The 2D-EMD $\rho^{2d}(p_x,p_y)$ was
calculated as
\begin{equation}
\rho^{2d}(p_x,p_y)= \int \rho(p_x,p_y,p_z)
 dp_z
\end{equation}
while the Compton profile $J(p_z)$ is given by
\begin{equation}
J(p_z) = \int \int \rho(p_x,p_y,p_z)
dp_xdp_y.
\end{equation}

\section{Results and Discussions}

In Fig.~\ref{fig:EMD}(a), we show the LDA band structure of LaOFeAs.
For $x$=0, three bands cross the $E_F$ around the $\Gamma$ point,
forming hole-like FSs [marked by green dots in (c)]
while two bands cross $E_F$ around $M$($\pi$,$\pi$),
forming electron-like FSs [marked by blue dots in (c)].
As electrons are added, the $\Gamma$ centered FSs shrink and
completely disappear around $x$=0.13.
The bands near $E_F$ are dominated by the Fe $d$ orbitals.
The FeAs layers are separated by insulating LaO layers,
with the result that the dispersion of these bands along $\Gamma-Z$
is small and, apart from a small $\Gamma$-centered 3D hole pocket, the
FSs are quasi two-dimensional.
Based on the fully three-dimensional computations,
we take advantage of this quasi two-dimensionality
and investigate quantities in the $k_x-k_y$ plane by integrating
out the $k_z$ component.

Figure \ref{fig:EMD}(b) shows a map of the theoretical
2D-EMD\cite{matsumoto2001}. This distribution can be described by an
inverted bell shape with fourfold symmetry. The peak is at the zone center
with tails extending over several unit cells. The dense contours
around high
symmetry points are signatures of the FS discontinuities. All these
features are hidden behind the large inverted bell shaped signal. In order
to investigate the Fermi surface topology in detail, we have employed both
the 2D Lock-Crisp-West (LCW) folding\cite{lcw1973,matsumoto2001} and the
2D-EMD anisotropy.

The 2D-LCW folding of the projected 2D-EMD $\rho^{2d}(p_x,p_y)$ is
defined by
\begin{equation}
n(k_x,k_y)=\sum_{G_x,G_y}\rho^{2d}(k_x+G_x,k_y+G_y),
\end{equation}
where $n(k_x,k_y)$ gives the number of occupied states at the point
$(k_x,k_y)$ in the first Brillouin zone by summing over all projected
reciprocal lattice vectors $(G_x,G_y)$. The effect of the matrix element
can be eliminated via the 2D-LCW folding process of Eq. (3), which thus
provides a tool for focusing on the FS features. The theoretical 2D-LCW
folding shown in Fig.~\ref{fig:EMD}(c) has been smoothed using a Gaussian
function with $\Delta p=0.17$ a.u., which is
typical of the resolutions available in high resolution Compton
scattering experiments. The positions and sizes of the FS pockets of the
undoped parent compound LaOFeAs found by our KKR band calculation are
shown as green dots for hole pockets and as blue dots for electron
pockets. Before the application of the aforementioned Gaussian broadening,
$n(k_x,k_y)$ shows a maximum $n_{max}$=29.4 and a minimum $n_{min}$=24.5.
The difference $n_{max}-n_{min}$=4.9 is consistent with five bands
crossing the Fermi level in the LDA calculation. Even after including
experimental resolution, the FS features are still quite visible as seen
in Fig.~\ref{fig:EMD}(c). The maximum of $n(k_x,k_y)$ at $M(\pi,\pi)$ is
associated with the electron pockets; the minimum at $\Gamma$ ($0$,$0$) is
related to the hole pockets.

\begin{figure}
\includegraphics[width=\hsize]{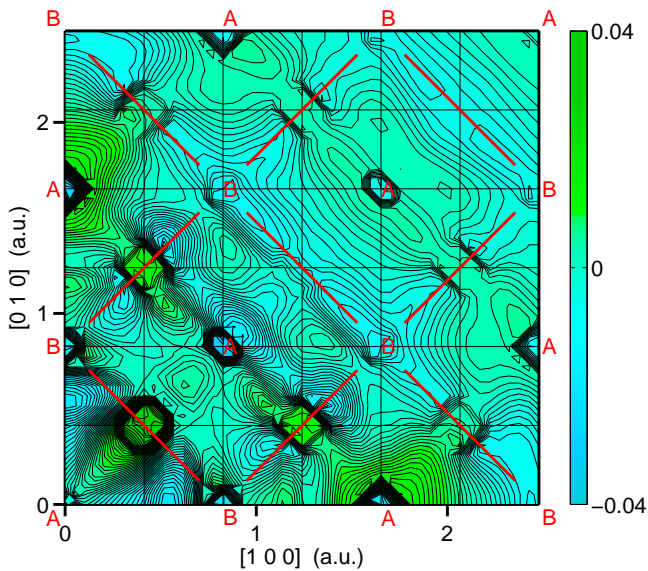}
\caption{(Color online)
Calculated 2D-EMD anisotropy in the parent compound LaOFeAs. Letters A and
B label $\Gamma$ points for odd and even $k$-space sublattices of the
hole-pockets as discussed in the text. The red lines are centered at $M$
points. The rapid changes in the momentum density along these lines are FS
signatures of the electron-pockets.
}\label{fig:alternating}
\end{figure}

Figure ~\ref{fig:alternating} shows the 2D-EMD anisotropy, found by
subtracting a smooth isotropic function from the 2D-EMD. FS features show
up as closely spaced contours around $\Gamma$($0$,$0$) and
$M$($\pi$,$\pi$) points. The momentum density around the $\Gamma(or M)$
points in the higher zones is seen to be lower (or higher) than the
average due to the presence of hole pockets (or electron pockets). The
zone-to-zone variation of intensity of these features can be understood
as a matrix element effect associated with the symmetry of the hole
pockets
at $\Gamma$ and electron pockets at $M$. For instance, the weak signal at
the origin ($0$,$0$) can be understood since the bands crossing the Fermi
level are predominantly $d$ orbitals, whereas only an $s$ orbital yields a
significant contribution to the momentum density at the origin. Owing to
interference effects, the FS features display a marked modulation from
zone to zone. The Fe atoms in the unit cell are located at high symmetry
positions, Fe1 (0,0,0) and Fe2 (0.5,0.5,0) (in units of lattice
constants). The wavefunctions of these two Fe atoms show a constructive
and destructive interference in momentum space, which can be represented
by the structure factor $S_{\bf G}=1+e^{-i\pi(m+n)}$, where
${\bf G}=(m\hat{x}+n\hat{y})[2\pi/a]$ is a reciprocal lattice vector.
Whereas
$S_{\bf G}$ is largest when $(m+n)$ is even, $S_{\bf G}$ vanishes when
$(m+n)$ is odd.
Therefore, the FS features show the alternating pattern seen in
Fig.~\ref{fig:alternating}. For the FS hole-pockets centered at $\Gamma$,
the FS features at B for odd ($m+n$) are much weaker than those at A
for even
($m+n$). Deviations from this rule are an indication of hybridization of
the Fe orbitals with other orbitals. For the FS electron-pockets centered
at $M$, a similar pattern is found. In Fig.~\ref{fig:alternating} we show
red lines crossing the B sites. The change of the momentum density along
the direction in which $\Delta m = \Delta n$ is more rapid than changes
along a direction for which $\Delta m = -\Delta n$.

\begin{figure}[H]
\includegraphics[width=\hsize]{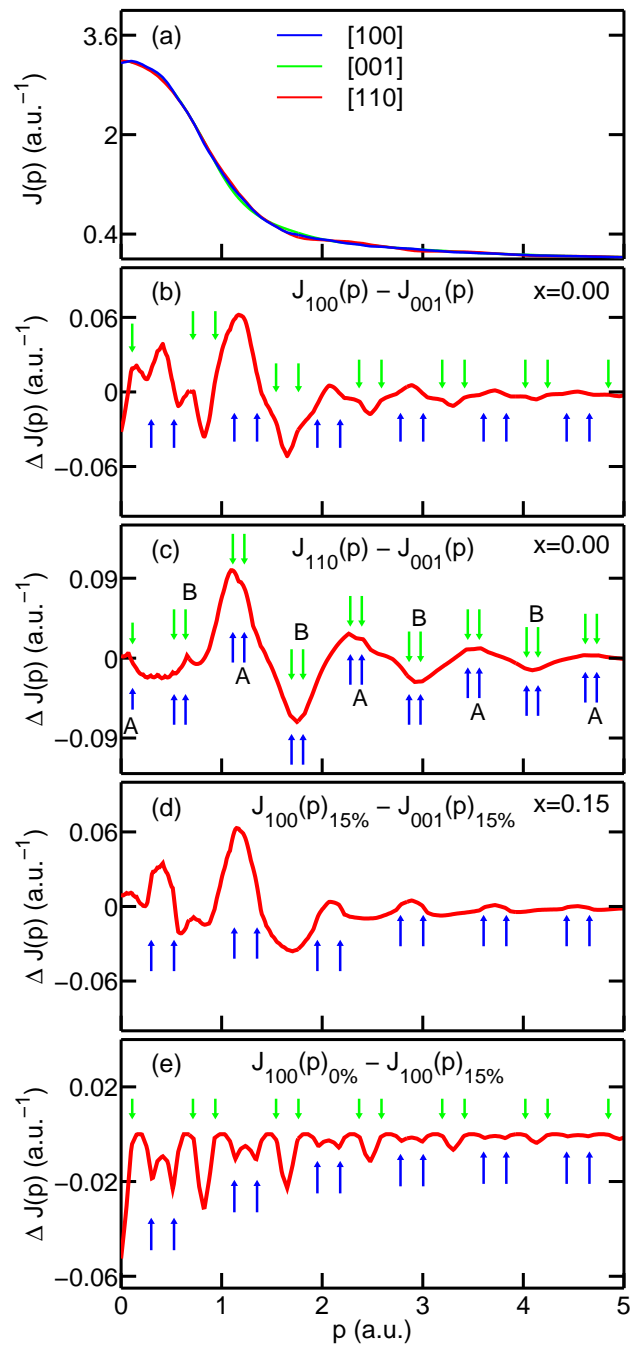}
\caption{(Color online)
(a) Directional Compton profiles(CPs) $J(p)$ along [100], [001] and [110]
for $x$=0. (b)-(e) Differences $\Delta J(p)$ between various pairs of CPs
for undoped ($x=0$) and doped ($x=0.15$) cases as indicated in the
figures.
Vertical arrows mark FS crossings with the electron
pockets centered at $M$ (blue) and the hole pockets at $\Gamma$ (green) as
discussed in the text.
}
\label{fig:cp}
\end{figure}

The theoretical CPs along [100], [110] and [001] are
shown in Fig.~\ref{fig:cp}(a). The CP along [001] is
a smooth curve, since there are no Fermi breaks along this direction.
Thus, we can use this profile for highlighting FS features from the
other profiles.

In Fig.~\ref{fig:cp}(b), we subtract the [001] from the [100] CP. The
resulting periodic patterns occur at the $\Gamma$ and $M$ points and are
associated with hole pockets (green arrows) and electron pockets (blue
arrows) respectively. The CP has a dip within the hole pocket regions and
a hump within the electron pocket regions. The FS breaks are clearly
visible and should be amenable to exploration via high resolution Compton
scattering experiments with statistics high enough to numerically
differentiate the difference profiles.

The same strategy is applied to the [110] CP in Fig.~\ref{fig:cp}(c);
however, the FS features are not as clear as in the [100] direction. The
main reason is that the contributions of the hole- and electron-like FSs
projected on [110] overlap each other and tend to cancel out.  The
interference pattern acts to amplify this effect as follows. When the EMD
is projected to form the CP, the projections of A and B are distinct
points along [110]. For hole pockets (centered at $\Gamma$), A~(B) has
strong~(weak) signals associated with FS breaks. For electron pockets
[centered at $(\pi,\pi)$], the red lines in the EMD are parallel to [110]
at the projection of A, while they are perpendicular to [110] at the
projection of B. We notice that the FS signals are strong only along the
red lines. Therefore, at the projection of A~(B), the FS signals are
strong~(weak) for both hole and electron pockets. In simple terms, at A,
strong hole pocket signals cancel strong electron pocket signals, while at
B weak hole pocket signals cancel weak electron pocket signals. Thus
[110] is not a suitable direction for studying the Fermi breaks.

Next, we discuss how FS breaks disappear with electron doping. The Fermi
level for $x$=0.15 shown in Fig.~1(a) at 0.075eV is obtained by a rigid
band shift. At this doping level, all hole pockets around $\Gamma$ are
removed. As indicated by blue arrows in Fig.~\ref{fig:cp}(d), the FS
breaks associated with the FS electron pockets remain in the [100] CP.
Compared to Fig.~\ref{fig:cp}(b), the dips associated with the hole
pockets have completely disappeared (green arrows). In
Fig.~\ref{fig:cp}(e), we subtract the [100] CP with $15$\% doping from the
[100] CP with $0$\% doping. An interesting pattern of periodic maxima and
minima appears around the $\Gamma$ and $M$ points which are identical with
the positions of the hole and electron pockets, respectively. This may
prove the most promising method of detecting FS signatures.

\section{Conclusions}

In conclusion, we have identified FS signatures in the momentum density of
LaOFeAs, finding alternating intensity patterns in the 2D-EMD due to the
symmetry of the crystal. FS signatures for both hole- and electron pockets
are shown to be relatively strong in the [100] CP in comparison to the
[110] CP. We thus conclude that the [100] direction is the favorable one
for studying FS signatures. Our analysis further indicates that a
doping-dependent experimental study should be able to determine at which
doping level the hole-like FS's disappear. The present work sets a
baseline for future experimental Compton scattering studies in the
pnictides.

\
\

\section{Acknowledgments}

This work is supported by the US Department of Energy, Office of Science,
Basic Energy Sciences contract DE-FG02-07ER46352, and benefited from the
allocation of supercomputer time at NERSC and Northeastern University's
Advanced Scientific Computation Center (ASCC). It was also sponsored by
the Stichting Nationale Computer Faciliteiten (NCF) for the use of
supercomputer facilities, with financial support from the Nederlandse
Organisatie voor Wetenschappelijk Onderzoek (Netherlands Organization for
Scientific Research).

 \end{document}